# *Numerical simulations of convection heat transfer in porous media using a cascaded lattice Boltzmann method*

Xiang-Bo Feng[a,b], Qing Liu[c,*], Ya-Ling He[b]

[a]*Shaanxi Key Laboratory of Safety and Durability of Concrete, Xijing University, Xi'an 710123, China*
[b]*School of Energy and Power Engineering, Xi'an Jiaotong University, Xi'an 710049, China*
[c]*School of Resources Engineering, Xi'an University of Architecture and Technology, Xi'an 710055, China*

([*]Corresponding author: qingliu1983@stu.xjtu.edu.cn)

**Abstract**

Convection heat transfer in porous media is a universal phenomenon in nature, and it is also frequently encountered in scientific and engineering fields. An in-depth understanding of the fundamental mechanism of convection heat transfer in porous media requires efficient and powerful numerical tools. In this paper, a cascaded lattice Boltzmann (CLB) method for convection heat transfer in porous media at the representative elementary volume (REV) scale is presented. In the CLB method, the flow field is solved by an isothermal CLB model with the D2Q9 lattice based on the generalized non-Darcy model, while the temperature field is solved by a temperature-based CLB model with the D2Q5 lattice. The key point is to incorporate the influence of the porous media into the CLB method by introducing the porosity and heat capacity ratio into the shift matrices. The effectiveness and practicability of the present method are validated by numerical simulations of several heat transfer problems in porous media at the REV scale. It is shown that the present method for convection heat transfer in porous media is second-order accurate in space. Moreover, in comparison with the Bhatnagar-Gross-Krook lattice Boltzmann method, the present method has sufficient tunable parameters and possesses better numerical stability.

## 1. Introduction

Fluid flow and convection heat transfer in porous media have attracted considerable attention since they are frequently encountered in scientific and engineering fields, such as carbon dioxide sequestration, contaminant transport in groundwater, underground mining, crude oil extraction, geothermal energy systems, construction engineering, geophysics and so on [1-4]. Fluid flow and heat transfer problems in porous media are multiscale in nature, and they usually involve three different process scales [3,5]: the region scale, the representative elementary volume (REV) scale, and the pore scale. The concept of REV is critical to understand and predict the behaviors of the macroscale parameters (i.e., porosity, permeability) of the porous media with complex structures. The REV is commonly defined as the minimum element of the porous media at which a given parameter is independent of the size of the REV. The characteristic size of the REV is considerably smaller than that of the region scale, but much larger than that of the pore scale (i.e., an REV contains a sufficiently large number of pores). Over the past three decades or so, fluid flow and convection heat transfer in porous media at the REV scale have been extensively studied using continuum-based numerical methods (e.g., finite-volume method, finite-difference method, finite-element method). A comprehensive review of the related subject can be found in the book by Nield and Bejan [3].

The lattice Boltzmann (LB) method [6-10], as a mesoscopic numerical method evolved from the lattice-gas automata (LGA) method [11], has been developed into a powerful and versatile numerical methodology for computational fluid dynamics and heat transfer [12-21]. Historically, the LB method evolved from the LGA method, quickly it was found that the LB equation can be rigorously derived from the continuous Boltzmann Bhatnagar-Gross-Krook (BGK) equation of the single-particle distribution function in the kinetic theory [22], which greatly solidifies the physical basis of the LB method. With its roots in LGA concept and the kinetic theory, the LB method has some distinctive features such as ideal for parallel computing, local nature of the operations, simple formulation and easy implementation of complex boundary conditions [12,13]. As a particle-based numerical method, the LB method principally aims to predict macroscopic quantities. Owing to its underlying kinetic nature, the LB method is particularly useful in studying fluid flow and transport phenomena involving interfacial dynamics and complex boundaries (e.g., fluid flow in porous media). Since the emergence of the LB method in 1988, its application in fluid flow and convection heat transfer in porous media has always been one of the most important themes of the method. In 1989, Succi et al. [23] first adopted the LB method to calculate the permeability of low-Reynolds-number flow in a three-dimensional random medium at the pore scale, and the Darcy's law was demonstrated. Since then fluid flow and convection heat transfer in porous media have been studied numerically by many scholars using the LB method. Generally, the LB models for fluid flow and convection heat transfer in porous media can be classified into two categories: the pore-scale method [23-26] and the REV-scale method [27-40]. In the pore-scale method, fluid flow and heat transfer in the pore spaces are directly modeled via the standard LB method, and the interactions between fluid and solid at the interface can be easily handled with the no-slip bounce-back rule. Pore-scale method offers unique opportunities to understand the underlying processes occurring in the complex porous structure. The detailed local information of the flow and heat transfer in the pore spaces can be obtained via pore-scale modeling, which can be used for evaluate constitutive closure relations.

In the REV-scale method, the porous medium is treated as continuous medium (i.e., the detailed geometric information of the medium is ignored), and the influence of the porous medium is considered by adding additional term to the LB equation (of the flow field) based on some semi-empirical models

(e.g., the Darcy model [31], the Brinkman-extended Darcy model [27-30] and the generalized non-Darcy model [32-40]). In this method, the statistical properties (e.g., porosity, permeability, effective thermal conductivity) are incorporated into the LB equations directedly without considering the detailed geometric structure of the medium. The REV-scale method is computationally efficient and can pass information into calculations at the region scale. Obviously, the accuracy of the REV-scale method depends on the semi-empirical models. However, based on appropriate semi-empirical models, the REV-scale method can produce reasonable results and it can be used for porous flow systems with large size. Moreover, the REV-scale method provides an ideal scale-bridging numerical tool to build multiscale method for multiscale simulations of fluid flow and heat transfer in porous media. After more than two decades of development, the REV-scale method has been developed into a powerful numerical tool for simulating fluid flow [27-33] and heat transfer [34-40] in porous media, providing some scientific guidelines and valuable references for practical applications.

In the LB community of fluid flow and heat transfer in porous media at the REV scale, the Bhatnagar-Gross-Krook (BGK) collision operator is still the most frequently used operator. Nevertheless, commonly, the BGK-LB model usually suffers from severe numerical instability at high Reynolds and Rayleigh numbers (the relaxation times are close to 0.5). In addition, the BGK collision operator does not have sufficient free parameters to describe anisotropic diffusion. Hence, several multiple-relaxation-time (MRT) LB models for simulating fluid flow and heat transfer in porous media have been proposed [37,39,40]. In the literature [30], a two-relaxation-time (TRT) LB model for simulating fluid flow in heterogeneous porous media based on the Brinkman-extended Darcy model has also been proposed. It has been well demonstrated that the MRT-LB and TRT-LB models are superior to their BGK counterparts in terms of physical principle, numerical accuracy and stability. In

2006, a cascaded collision operator was proposed by Geier et al. [41]. In the cascaded collision operator, the collision process is performed in terms of central moments in a moving frame of reference, while the collision process of the MRT collision operator is performed in terms of raw moments in a rest frame of reference (the central moments are obtained by shifting the particle velocity by the local fluid velocity, and the moments without such shift are called the raw moments) [42,43]. Thus the cascaded collision operator is also interpreted as ''central-moments-based'' collision operator [43,44]. As compared with the BGK-LB method, the cascaded lattice Boltzmann (CLB) method can significantly enhance the numerical stability [41,43,45]. Moreover, the CLB method shows some advantages over the classical MRT-LB method in terms of Galilean invariance and numerical stability [41,43,45]. Most recently, Fei et al. [45] proposed a CLB method for incompressible thermal flows. Fei et al.'s method has several distinctive features: (1) a consistent forcing scheme is employed to incorporate the forcing effect (see Ref. [46] for details); (2) the implementation processes are greatly simplified by using simplified raw-moment sets; (3) as compared with several existing MRT-LB models [47,48] for the temperature equation, Fei et al.'s method shows better Galilean invariance. Hence, in this paper we aim to develop an REV-scale LB method for convection heat transfer in porous media in the framework of the cascaded collision operator, which can be viewed as an extension to some previous works [32,34,37,45]. It is expected that the proposed CLB method can serve as an efficient and powerful numerical tool for studying convection heat transfer in porous media. The rest of this paper is organized as follows. In Section 2, the CLB method for convection heat transfer in porous media at the REV scale is presented in detail. In Section 3, the effectiveness and practicability of the REV-scale CLB method are validated by numerical simulations of several heat transfer problems in porous media. Finally, some conclusions are made in Section 4.

## 2. CLB method for convection heat transfer in porous media

For fluid flow and convection heat transfer in isotropic and rigid porous media, based on the generalized non-Darcy model, the macroscopic governing equations under local thermal equilibrium condition are given by [3,34,49]

$$\nabla \cdot \mathbf{u} = 0 \tag{1}$$

$$\frac{\partial \mathbf{u}}{\partial t} + (\mathbf{u} \cdot \nabla)\left(\frac{\mathbf{u}}{\phi}\right) = -\frac{1}{\rho_0}\nabla(\phi p) + v_e \nabla^2 \mathbf{u} + \mathbf{F} \tag{2}$$

$$\frac{\partial(\sigma T)}{\partial t} + \mathbf{u} \cdot \nabla T = \nabla \cdot (\alpha_e \nabla T) \tag{3}$$

where $\mathbf{u}$, $p$, and $T$ are the volume-averaged fluid velocity, pressure, and temperature, respectively, $\rho_0$ is the reference fluid density, $\phi$ is the porosity, $v_e$ is the effective kinematic viscosity, $\alpha_e$ is the effective thermal diffusivity, and $\sigma = \left[\phi \rho_f c_{pf} + (1-\phi)\rho_m c_{pm}\right]/(\rho_f c_{pf})$ is the heat capacity ratio ($c_p$ is the specific heat, and the subscripts $f$ and $m$ denote the fluid and solid matrix, respectively). $\mathbf{F} = (F_x, F_y)$ represents the total body force induced by the porous matrix and external force, which can be expressed as [32,49,50]

$$\mathbf{F} = -\frac{\phi v}{K}\mathbf{u} - \frac{\phi F_\phi}{\sqrt{K}}|\mathbf{u}|\mathbf{u} + \phi \mathbf{G} \tag{4}$$

where $v$ is the kinematic viscosity of the fluid ($v$ is not necessarily the same as $v_e$), $K$ is the permeability of the porous medium, $F_\phi$ is the structure function or inertia coefficient, and $|\mathbf{u}| = \sqrt{u_x^2 + u_y^2}$, in which $u_x$ and $u_y$ are the $x$- and $y$-component of the fluid velocity $\mathbf{u}$, respectively. The first and second terms on the right hand side of Eq. (4) are the linear (Darcy's term) and nonlinear (Forchheimer's term) drag forces of the porous matrix, respectively. Without the nonlinear drag force term ($F_\phi = 0$), Eq. (2) reduces to the Brinkman-extended Darcy equation [3,32]. According to the Boussinesq approximation, the body force $\mathbf{G}$ is given by $\mathbf{G} = -\mathbf{g}\beta(T - T_0) + \mathbf{a}$, where $\mathbf{g}$ is the gravitational acceleration, $\beta$ is the thermal expansion coefficient, $T_0$ is the reference temperature, and **a** is the acceleration induced by other external force fields.

The inertia coefficient $F_\phi$ depends on the geometry of the medium. Based on Ergun's experimental investigations [51], the inertia coefficient $F_\phi$ and the permeability $K$ can be expressed as [52]

$$F_\phi = \frac{1.75}{\sqrt{150\phi^3}}, \quad K = \frac{\phi^3 d_p^2}{150(1-\phi)^2} \tag{5}$$

where $d_p$ is the solid particle diameter.

*2.1 D2Q9 CLB model for the flow field*

In this subsection, the CLB model for the flow field (governed by Eqs. (1) and (2)) is presented. The D2Q9 lattice [10] is employed, where the nine discrete velocities $\{\mathbf{e}_i | i = 0, 1, \ldots, 8\}$ are given by

$$\mathbf{e}_i = \begin{cases} (0,0), & i = 0 \\ \left(\cos\left[(i-1)\pi/2\right], \sin\left[(i-1)\pi/2\right]\right)c, & i = 1 \sim 4 \\ \left(\cos\left[(2i-9)\pi/4\right], \sin\left[(2i-9)\pi/4\right]\right)\sqrt{2}c, & i = 5 \sim 8 \end{cases} \tag{6}$$

in which $c = \delta_x/\delta_t$ is the lattice speed with $\delta_t$ and $\delta_x$ being the discrete time step and the lattice spacing (in the x-direction), respectively. In this work, the lattice speed $c$ is set to 1 ($\delta_x = \delta_t$) such that all the relevant quantities are dimensionless.

The raw moments $\{k_{mn}\}$ and central moments $\{\tilde{k}_{mn}\}$ of the discrete density distribution function $f_i$ are defined as follows [41]

$$k_{mn} = \left\langle f_i \middle| e_{ix}^m e_{iy}^n \right\rangle \tag{7}$$

$$\tilde{k}_{mn} = \left\langle f_i \middle| (e_{ix} - u_x)^m (e_{iy} - u_y)^n \right\rangle \tag{8}$$

The equilibrium values $\{k_{mn}^{eq}\}$ and $\{\tilde{k}_{mn}^{eq}\}$ can be defined analogously by replacing $f_i$ with the discrete equilibrium density distribution function $f_i^{eq}$. In this work, the following simplified raw-moment and central-moment sets are employed [45,53]

$$|m_i\rangle = [k_{00}, k_{10}, k_{01}, k_{20}, k_{02}, k_{11}, k_{21}, k_{12}, k_{22}]^{\mathrm{T}} \qquad (9)$$

$$|\tilde{m}_i\rangle = [\tilde{k}_{00}, \tilde{k}_{10}, \tilde{k}_{01}, \tilde{k}_{20}, \tilde{k}_{02}, \tilde{k}_{11}, \tilde{k}_{21}, \tilde{k}_{12}, \tilde{k}_{22}]^{\mathrm{T}} \qquad (10)$$

The CLB equation for the flow field consists of two processes: the collision process and streaming process. The collision process is executed in the central-moment space [45,46]

$$|\tilde{m}_i^*\rangle = |\tilde{m}_i\rangle - \Lambda\left(|\tilde{m}_i\rangle - |\tilde{m}_i^{eq}\rangle\right) + \delta_t\left(\mathbf{I} - \frac{\Lambda}{2}\right)|\tilde{S}_i\rangle \qquad (11)$$

where $\{\tilde{m}_i^{eq}\}$ are the equilibrium central moments, $\{\tilde{S}_i\}$ are forcing terms in the central-moment space, and $\Lambda$ is a block-diagonal relaxation matrix. Here $\{\tilde{m}_i^*\}$ are the post-collision central moments.

The relationship between the central moments and raw moments can be expressed as [45]

$$|\tilde{m}_i\rangle = \mathbf{N}|m_i\rangle, \quad |m_i\rangle = \mathbf{M}|f_i\rangle \qquad (12)$$

where $\mathbf{N}$ is the shift matrix, and $\mathbf{M}$ is the transformation matrix. As defined in Eq. (12), the raw moments are transformed from $f_i$ through the transformation matrix $\mathbf{M}$, and the central moments are shifted from the raw moments through the shift matrix $\mathbf{N}$. The transformation matrix $\mathbf{M}$ is given by [45]

$$\mathbf{M} = \begin{bmatrix} 1 & 1 & 1 & 1 & 1 & 1 & 1 & 1 & 1 \\ 0 & 1 & 0 & -1 & 0 & 1 & -1 & -1 & 1 \\ 0 & 0 & 1 & 0 & -1 & 1 & 1 & -1 & -1 \\ 0 & 1 & 0 & 1 & 0 & 1 & 1 & 1 & 1 \\ 0 & 0 & 1 & 0 & 1 & 1 & 1 & 1 & 1 \\ 0 & 0 & 0 & 0 & 0 & 1 & -1 & 1 & -1 \\ 0 & 0 & 0 & 0 & 0 & 1 & 1 & -1 & -1 \\ 0 & 0 & 0 & 0 & 0 & 1 & -1 & -1 & 1 \\ 0 & 0 & 0 & 0 & 0 & 1 & 1 & 1 & 1 \end{bmatrix} \qquad (13)$$

To include the effect of the porous medium, the shift matrix $\mathbf{N}$ is given by

$$\mathbf{N} = \begin{bmatrix} 1 & 0 & 0 & 0 & 0 & 0 & 0 & 0 & 0 \\ -u_x & 1 & 0 & 0 & 0 & 0 & 0 & 0 & 0 \\ -u_y & 0 & 1 & 0 & 0 & 0 & 0 & 0 & 0 \\ u_x^2/\phi & -2u_x/\phi & 0 & 1 & 0 & 0 & 0 & 0 & 0 \\ u_y^2/\phi & 0 & -2u_y/\phi & 0 & 1 & 0 & 0 & 0 & 0 \\ u_x u_y/\phi & -u_y/\phi & -u_x/\phi & 0 & 0 & 1 & 0 & 0 & 0 \\ -u_x^2 u_y/\phi & 2u_x u_y/\phi & u_x^2/\phi & -u_y & 0 & -2u_x & 1 & 0 & 0 \\ -u_y^2 u_x/\phi & u_y^2/\phi & 2u_x u_y/\phi & 0 & -u_x & -2u_y & 0 & 1 & 0 \\ u_x^2 u_y^2/\phi^2 & -2u_x u_y^2/\phi^2 & -2u_y u_x^2/\phi^2 & u_y^2/\phi & u_x^2/\phi & 4u_x u_y/\phi & -2u_y/\phi & -2u_x/\phi & 1 \end{bmatrix} \quad (14)$$

The block-diagonal relaxation matrix $\mathbf{\Lambda}$ is given by [45,53]

$$\mathbf{\Lambda} = \mathrm{diag}\left([s_0, s_1, s_1], \begin{bmatrix} s_+ & s_- \\ s_- & s_+ \end{bmatrix}, [s_v, s_3, s_3, s_4]\right) \quad (15)$$

where $s_+ = (s_b + s_v)/2$ and $s_- = (s_b - s_v)/2$. The effective kinematic viscosity and the bulk viscosity are related with the relaxation parameters as $v_e = c_s^2 (s_v^{-1} - 0.5)\delta_t$ and $v_b = c_s^2 (s_b^{-1} - 0.5)\delta_t$, respectively, where $c_s = c/\sqrt{3}$ is the sound speed of the D2Q9 model.

The equilibrium central moments $\{\tilde{m}_i^{eq}\}$ and the forcing terms $\{\tilde{S}_i\}$ in the central-moment space are given by

$$|\tilde{m}_i^{eq}\rangle = [\rho, 0, 0, \rho c_s^2, \rho c_s^2, 0, 0, 0, \rho c_s^4]^{\mathrm{T}} \quad (16)$$

$$|\tilde{S}_i\rangle = [0, \rho F_x, \rho F_y, 0, 0, 0, c_s^2 \rho F_y, c_s^2 \rho F_x, 0]^{\mathrm{T}} \quad (17)$$

The streaming process is still executed in the velocity space

$$f_i(\mathbf{x} + \mathbf{e}_i \delta_t, t + \delta_t) = f_i^*(\mathbf{x}, t) \quad (18)$$

where $f_i^*$ is the post-collision discrete density distribution function determined by $|f_i^*\rangle = \mathbf{M}^{-1}\mathbf{N}^{-1}|\tilde{m}_i^*\rangle$.

The equilibrium raw moments $\{m_i^{eq}\}$ and the forcing terms $\{S_i\}$ in the raw-moment space are given by

$$\left| m_i^{eq} \right\rangle = \begin{bmatrix} \rho \\ \rho u_x \\ \rho u_y \\ \dfrac{\rho}{3} + \dfrac{\rho u_x^2}{\phi} \\ \dfrac{\rho}{3} + \dfrac{\rho u_y^2}{\phi} \\ \dfrac{\rho u_x u_y}{\phi} \\ \dfrac{\rho u_y}{3} + \dfrac{\rho u_x^2 u_y}{\phi} \\ \dfrac{\rho u_x}{3} + \dfrac{\rho u_x u_y^2}{\phi} \\ \dfrac{\rho}{9} + \dfrac{\rho(u_x^2 + u_y^2)}{3\phi} + \dfrac{\rho u_x^2 u_y^2}{\phi^2} \end{bmatrix}, \quad \left| S_i \right\rangle = \begin{bmatrix} 0 \\ \rho F_x \\ \rho F_y \\ \dfrac{2\rho u_x F_x}{\phi} \\ \dfrac{2\rho u_y F_y}{\phi} \\ \dfrac{\rho(u_x F_y + u_y F_x)}{\phi} \\ \dfrac{\rho F_y}{3} + \dfrac{\rho F_y u_x^2}{\phi} + 2\dfrac{\rho F_x u_x u_y}{\phi} \\ \dfrac{\rho F_x}{3} + \dfrac{\rho F_x u_y^2}{\phi} + 2\dfrac{\rho F_y u_y u_x}{\phi} \\ \dfrac{2}{3}\dfrac{\rho(u_x F_x + u_y F_y)}{\phi} + 2\dfrac{\rho F_x u_x u_y^2}{\phi^2} + 2\dfrac{\rho F_y u_y u_x^2}{\phi^2} \end{bmatrix} \quad (19)$$

The discrete equilibrium density distribution function $f_i^{eq}$ in the velocity space can be obtained via $\left| f_i^{eq} \right\rangle = \mathbf{M}^{-1} \left| m_i^{eq} \right\rangle$.

The density $\rho$ and velocity $\mathbf{u}$ are defined by

$$\rho = \sum_{i=0}^{8} f_i \quad (20)$$

$$\rho \mathbf{u} = \sum_{i=0}^{8} \mathbf{e}_i f_i + \dfrac{\delta_t}{2} \rho \mathbf{F} \quad (21)$$

The fluid pressure $p$ is defined as $p = \rho c_s^2 / \phi$. Eq. (21) is a nonlinear equation for the velocity $\mathbf{u}$. By introducing a temporal velocity $\mathbf{v}$, the macroscopic fluid velocity $\mathbf{u}$ can be calculated explicitly by [32]

$$\mathbf{u} = \dfrac{\mathbf{v}}{l_0 + \sqrt{l_0^2 + l_1 |\mathbf{v}|}} \quad (22)$$

where

$$\mathbf{v} = \dfrac{1}{\rho} \sum_{i=0}^{8} \mathbf{e}_i f_i + \dfrac{\delta_t}{2} \phi \mathbf{G}, \quad l_0 = \dfrac{1}{2}\left(1 + \phi \dfrac{\delta_t}{2} \dfrac{v}{K}\right), \quad l_1 = \phi \dfrac{\delta_t}{2} \dfrac{F_\phi}{\sqrt{K}} \quad (23)$$

### 2.2 D2Q5 CLB model for the temperature field

In this subsection, the CLB model for the temperature field (governed by Eq. (3)) is presented. The D2Q5 lattice is employed, where the five discrete velocities $\{\mathbf{e}_i | i = 0, \ldots, 4\}$ can be found in Eq.

(6). Similarly, the raw moments $\{k_{mn}^T\}$ and central moments $\{\tilde{k}_{mn}^T\}$ of the discrete temperature distribution function $g_i$ are defined as follows [45]

$$k_{mn}^T = \langle g_i | e_{ix}^m e_{iy}^n \rangle \tag{24}$$

$$\tilde{k}_{mn}^T = \langle g_i | (e_{ix} - u_x)^m (e_{iy} - u_y)^n \rangle \tag{25}$$

For the temperature field, the following simplified raw-moment and central-moment sets are employed [45]

$$|n_i\rangle = \left[ k_{00}^T, k_{10}^T, k_{01}^T, k_{20}^T, k_{02}^T \right]^\top \tag{26}$$

$$|\tilde{n}_i\rangle = \left[ \tilde{k}_{00}^T, \tilde{k}_{10}^T, \tilde{k}_{01}^T, \tilde{k}_{20}^T, \tilde{k}_{02}^T \right]^\top \tag{27}$$

The CLB equation for the temperature field also consists of two processes, i.e., the collision process and streaming process. The collision process executed in the central-moment space is given by

$$|\tilde{n}_i^*\rangle = |\tilde{n}_i\rangle - \mathbf{Q}\left( |\tilde{n}_i\rangle - |\tilde{n}_i^{eq}\rangle \right) \tag{28}$$

where $\{\tilde{n}_i^{eq}\}$ are the equilibrium central moments, and $\mathbf{Q} = \text{diag}(\zeta_T, \zeta_\alpha, \zeta_\alpha, \zeta_e, \zeta_\nu)$ is the relaxation matrix. Here $\{\tilde{n}_i^*\}$ are the post-collision central moments.

The relationship between the central moments and raw moments is [45]

$$|\tilde{n}_i\rangle = \mathbf{N}_T |n_i\rangle, \quad |n_i\rangle = \mathbf{M}_T |g_i\rangle \tag{29}$$

where $\mathbf{N}_T$ is the shift matrix, and $\mathbf{M}_T$ is the transformation matrix. The raw moments are transformed from $g_i$ through the transformation matrix $\mathbf{M}_T$, and the central moments are shifted from the raw moments through the shift matrix $\mathbf{N}_T$. The transformation matrix $\mathbf{M}_T$ is given by [45]

$$\mathbf{M}_T = \begin{bmatrix} 1 & 1 & 1 & 1 & 1 \\ 0 & 1 & 0 & -1 & 0 \\ 0 & 0 & 1 & 0 & -1 \\ 0 & 1 & 0 & 1 & 0 \\ 0 & 0 & 1 & 0 & 1 \end{bmatrix} \tag{30}$$

To consider the effect of the heat capacity ratio, the shift matrix $\mathbf{N}_T$ is given by

$$\mathbf{N}_T = \begin{bmatrix} 1 & 0 & 0 & 0 & 0 \\ -u_x/\sigma & 1 & 0 & 0 & 0 \\ -u_y/\sigma & 0 & 1 & 0 & 0 \\ u_x^2/\sigma & -2u_x & 0 & 1 & 0 \\ u_y^2/\sigma & 0 & -2u_y & 0 & 1 \end{bmatrix} \tag{31}$$

The equilibrium central moments $\{\tilde{n}_i^{eq}\}$ and raw moments $\{n_i^{eq}\}$ are given by

$$\left| \tilde{n}_i^{eq} \right\rangle = \left[ \sigma T, 0, 0, c_{sT}^2 T, c_{sT}^2 T \right]^{\mathrm{T}} \tag{32}$$

$$\left| n_i^{eq} \right\rangle = \left[ \sigma T, u_x T, u_y T, \left(c_{sT}^2 + u_x^2\right)T, \left(c_{sT}^2 + u_y^2\right)T \right]^{\mathrm{T}} \tag{33}$$

where $c_{sT}$ is the sound speed of the D2Q5 model.

The discrete equilibrium temperature distribution function $g_i^{eq}$ in the velocity space can be obtained via $\left| g_i^{eq} \right\rangle = \mathbf{M}_T^{-1} \left| n_i^{eq} \right\rangle$. Explicitly, $g_i^{eq}$ is given by

$$\begin{aligned}
g_0^{eq} &= \sigma T - \varpi T - \left(u_x^2 + u_y^2\right)T \\
g_1^{eq} &= \frac{\varpi}{4}T + \frac{1}{2}u_x T + \frac{1}{2}u_x^2 T \\
g_2^{eq} &= \frac{\varpi}{4}T + \frac{1}{2}u_y T + \frac{1}{2}u_y^2 T \\
g_3^{eq} &= \frac{\varpi}{4}T - \frac{1}{2}u_x T + \frac{1}{2}u_x^2 T \\
g_4^{eq} &= \frac{\varpi}{4}T - \frac{1}{2}u_y T + \frac{1}{2}u_y^2 T
\end{aligned} \tag{34}$$

where $\varpi \in (0,1)$ is a parameter of the model ($c_{sT} = c\sqrt{\varpi/2} = \sqrt{\varpi/2}$).

The streaming process is executed in the velocity space

$$g_i\left(\mathbf{x} + \mathbf{e}_i \delta_t, t + \delta_t\right) = g_i^*\left(\mathbf{x}, t\right) \tag{35}$$

where $g_i^*$ is the post-collision discrete temperature distribution function determined by $\left| g_i^* \right\rangle = \mathbf{M}_T^{-1} \mathbf{N}_T^{-1} \left| \tilde{n}_i^* \right\rangle$.

The temperature $T$ is computed by

$$\sigma T = \sum_{i=0}^{4} g_i \tag{36}$$

The effective thermal diffusivity $\alpha_e$ is defined as

$$\alpha_e = c_{sT}^2 \left( \frac{1}{\zeta_\alpha} - \frac{1}{2} \right) \delta_t \tag{37}$$

In this section, the CLB method for convection heat transfer in porous media at the REV scale is presented. Through the Chapman-Enskog analysis [42,45], the macroscopic governing equations (1)-(3) can be reproduced from the CLB method in the low-Mach-number limit. When the shift matrices $\mathbf{N}$ and $\mathbf{N}_T$ are unit matrices, the present CLB method degrades into the non-orthogonal MRT-LB method in Ref. [54] with some high-order terms in the equilibrium raw moments and forcing terms. It is noted that as $\phi \rightarrow 1$, the present CLB method reduces to the CLB method [45] in the absence of porous media.

## 3. Numerical simulations

In this section, numerical simulations of mixed convection flow in a porous channel and natural convection in a porous cavity are carried out to demonstrate the practicability and accuracy of the proposed CLB method for convection heat transfer in porous media. In addition to the porosity $\phi$ and the heat capacity ratio $\sigma$, the flow and heat transfer processes are characterized by several dimensionless parameters: the Rayleigh number $Ra$, the Darcy number $Da$, the Reynolds number $Re$ (for mixed convection flow), the Prandtl number $Pr$, the viscosity ratio $J$, and the thermal diffusivity ratio $\Gamma$, which are defined as follows

$$Da = \frac{K}{L_c^2}, \quad Ra = \frac{g\beta \Delta T L_c^3}{\nu \alpha}, \quad Re = \frac{L_c u_c}{\nu}, \quad Pr = \frac{\nu}{\alpha}, \quad J = \frac{\nu_e}{\nu}, \quad \Gamma = \frac{\alpha_e}{\alpha} \tag{38}$$

where $L_c$ is the characteristic length, $u_c$ is the characteristic velocity, $\Delta T$ is the temperature difference (characteristic temperature), and $\alpha$ is the thermal diffusivity of the fluid.

In simulations, we set $\delta_x = \delta_y = \delta_t = 1$ ($c = 1$), $J = 1$, and $c_{sT} = \sqrt{1/8}$ ($\varpi = 1/4$). According to Refs. [45,47], the free relaxation rates are chosen as follows: $s_0 = s_1 = 1$, $\zeta_b = \zeta_v$, $s_3 = 1.2$, $s_4 = 1.8$, and $\zeta_T = \zeta_e = \zeta_v = 1$. Unless otherwise specified, the non-equilibrium extrapolation scheme [55] is employed to treat the velocity and temperature boundary conditions. All the simulations are performed on a personal computer (processor: Inter(R) Core(TM) i5-8400 CPU@2.80GHz; RAM: 8.0 GB).

### *3.1 Mixed convection flow in a porous channel*

In this subsection, we consider the mixed convection flow in a channel filled with porous media (see Fig. 1). The distance between the two parallel plates is $H$, the upper plate is hot ($T=T_h$) and moves along the *x*-direction with a uniform velocity $u_0$, while the cold bottom plate is static ($T=T_c$) and a constant normal flow of fluid is injected (with a uniform velocity $u_1$) through the bottom plate and is withdrawn at the same rate from the upper plate.

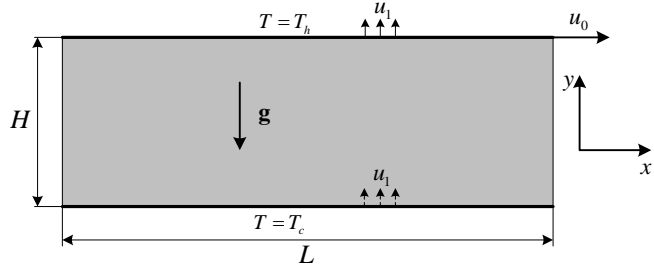

Fig. 1. Schematic diagram of mixed convection flow in a porous channel.

Without the nonlinear drag force ($F_\phi = 0$), the flow is governed by the following equations (at the steady state) [34]

$$\frac{u_y}{\phi}\frac{\partial u_x}{\partial y} = v_e \frac{\partial^2 u_x}{\partial y^2} - \frac{\phi v}{K} u_x \tag{39}$$

$$\frac{1}{\rho_0}\frac{\partial p}{\partial y} = g\beta(T-T_0) - \frac{v}{K}u_y + a_y \tag{40}$$

$$u_y \frac{\partial T}{\partial y} = \nabla \cdot (\alpha_e \nabla T) \tag{41}$$

where $T_0 = (T_h + T_c)/2$ is the reference temperature, and $a_y$ is the external force in the *y*-direction:

$$a_y = \frac{v}{K}u_1 - g\beta\Delta T \left[\frac{\exp(yu_1/\alpha_e)-1}{\exp(Hu_1/\alpha_e)-1}\right] \tag{42}$$

The analytical solutions of Eqs. (39)-(41) are given by [34]

$$u_x = u_0 \exp\left[\vartheta_1\left(\frac{y}{H}-1\right)\right]\frac{\sinh(\vartheta_2 \cdot y/H)}{\sinh(\vartheta_2)}, \quad u_y = u_1 \tag{43}$$

$$T = T_c + \Delta T \frac{\exp(Pr_e Re \cdot y/H) - 1}{\exp(Pr_e Re) - 1} \tag{44}$$

where $Re = Hu_1/v$ is the Reynolds number, $Pr_e = v/\alpha_e$ is the effective Prandtl number, $\Delta T = T_h - T_c$ is the temperature difference. The two parameters $\vartheta_1$ and $\vartheta_2$ in Eq. (43) are given by $\vartheta_1 = Re/(2\phi J)$ and $\vartheta_2 = \sqrt{Re^2 + 4\phi^3 J/Da}/(2\phi J)$. In simulations, we set $Ra = 100$, $\phi = 0.6$, $Pr_e = Pr = 1$, $Da = 0.01$, $J = \Gamma = \sigma = 1$, and $u_0 = u_1 = 0.01$. Periodic boundary conditions are imposed in the $x$-direction, and a grid size of $N_x \times N_y = 32 \times 32$ is adopted. The relaxation rates $s_v$ and $\zeta_\alpha$ are determined by $s_v^{-1} = 0.5 + JHu_1/(c_s^2 Re\delta_t)$ and $\zeta_\alpha^{-1} = 0.5 + \Gamma c_s^2 (s_v^{-1} - 0.5)/(Jc_{sT}^2 Pr)$, respectively. In Fig. 2, the normalized velocity and temperature profiles for different $Re$ with $Da = 0.01$ and $\phi = 0.6$ are plotted and compared with the analytical solutions. Here, $\theta = (T - T_c)/\Delta T$ is the normalized temperature. As shown in the figure, the present results agree well with the analytical ones.

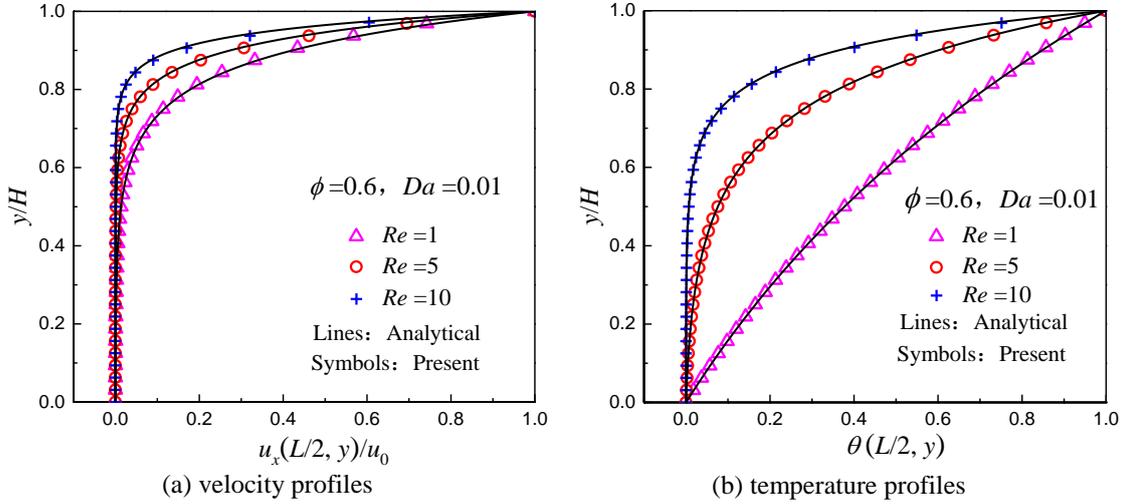

(a) velocity profiles  (b) temperature profiles

Fig. 2. Velocity and temperature profiles for different $Re$ with $Da = 0.01$ and $\phi = 0.6$.

Numerical simulations are carried out to evaluate the spatial accuracy of the present CLB method. In simulations, the relaxation rate $s_v$ is fixed at 1, $Re = 5$, $Da = 0.01$, and $\phi = 0.6$. The grid number $N_y$ varies from 16 to 96. The relative global error of a variable ($\mathbf{u}$ or $T$) is defined by

$$E(\Gamma) = \frac{\sqrt{\sum_{\mathbf{x}} |\Gamma_a(\mathbf{x}) - \Gamma_{lb}(\mathbf{x})|^2}}{\sqrt{\sum_{\mathbf{x}} |\Gamma_a(\mathbf{x})|^2}} \qquad (45)$$

where the summation $\sum_{\mathbf{x}}$ is over the entire domain, and $\Gamma_a$ and $\Gamma_{lb}$ represent the analytical and numerical solutions, respectively. The relative global errors of the flow and temperature fields are plotted logarithmically in Fig. 3, where the symbols denote the results measured by the CLB method and the solid lines represent the least-square fittings. As shown in the figure, the slopes of the fitting lines for the flow and temperature fields are 2.032 and 1.996, respectively. The second-order accurate in space of the present CLB method is clearly shown.

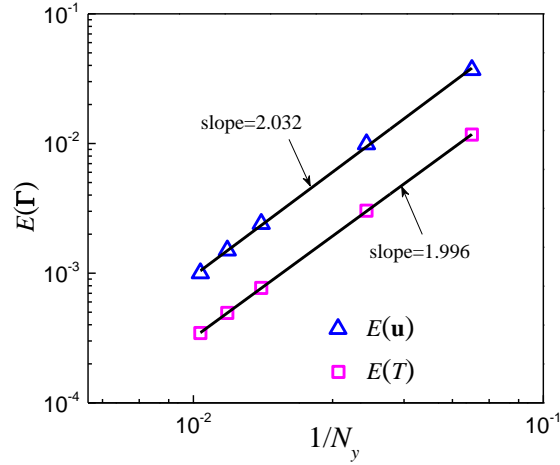

Fig. 3. Relative global errors of the flow and temperature fields with $Re = 5$, $Da = 0.01$, and $\phi = 0.6$.

As stated in the Introduction, the CLB method can significantly enhance the numerical stability as compared with the BGK-LB method [41,43,45]. In what follows, numerical simulaions are conducted to confirm that the present CLB method is more stable than the BGK-LB method [34] at low viscosities. In simulations, $Ra = 100$, $Re = 5$, $Pr = 1$, $\phi = 0.6$, $Da = 0.01$, $J = \Gamma = \sigma = 1$, and $N_x \times N_y = 32 \times 32$. In Fig. 4, the normalized velocity profiles at low viscosities ($v = 5 \times 10^{-4}$ and $1.0 \times 10^{-4}$) are shown. As shown in the figure, the BGK-LB method [34] is unstable at $v = 5 \times 10^{-4}$ ($\tau_v = 0.5015$) when $t = 2200\delta_t$ (see Fig. 4a), while the present CLB method ($s_v^{-1} = \tau_v = 0.5015$) is still stable (see Fig. 4b) and the steady-state result agrees well with the analytical result (see Fig. 4c).

This demonstrates that the present CLB method is indeed more stable that the BGK-LB method. To strength this statement, we further reduce the viscosity to a smaller value and find that the present CLB method still works well (see Fig. 4d).

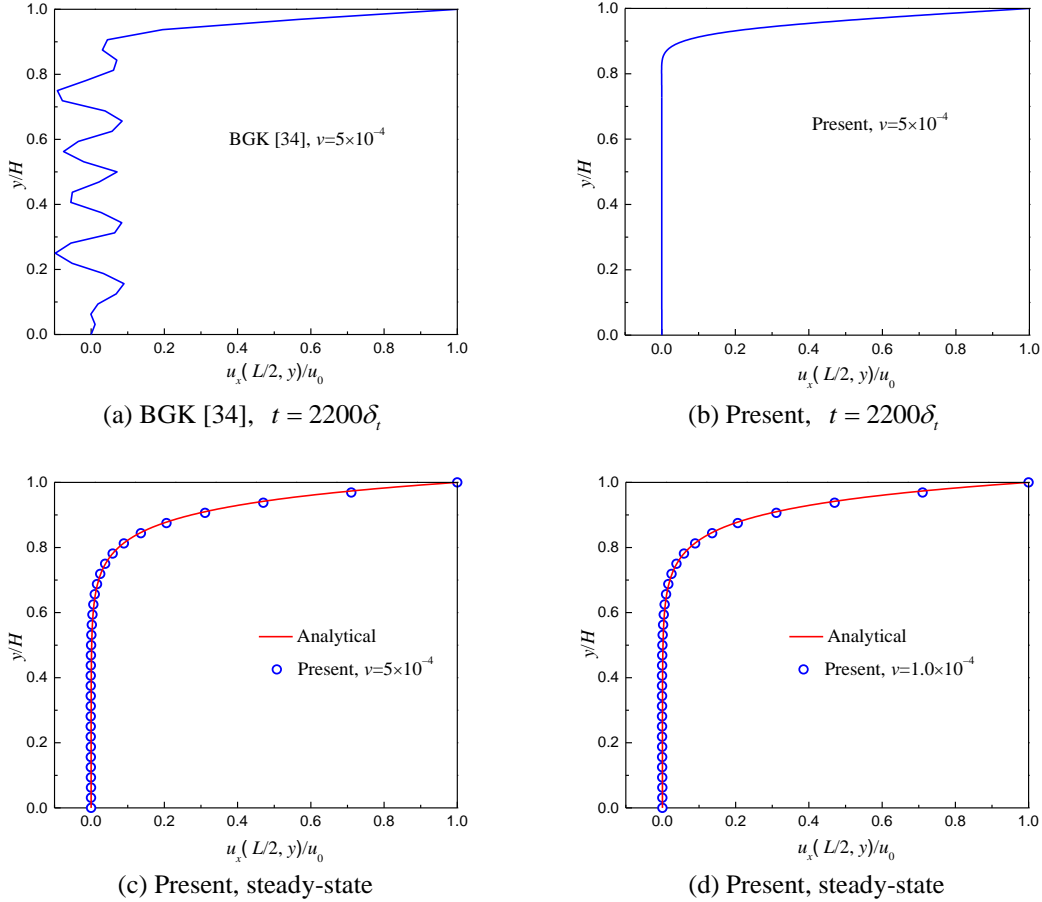

Fig. 4. Velocity profiles of the mixed convection flow at low viscosities ($v = 5\times10^{-4}$ and $1.0\times10^{-4}$).

*3.2 Natural convection in a porous cavity*

We now consider the natural convection in a fluid-saturated porous cavity (see Fig. 5), which has been studied by many researchers [34,37,38,49,54]. The left and right walls of the cavity are kept at constant temperatures $T_h$ and $T_c$ ($T_h > T_c$), respectively, while the top and bottom walls are adiabatic. The average Nusselt number $\overline{Nu}$ of the left (or right) wall is defined by $\overline{Nu} = \int_0^H Nu(y)\,dy/H$, where $Nu(y) = -L(\partial T/\partial x)_{wall}/\Delta T$ is the local Nusselt number, $\Delta T = T_h - T_c$ is the temperature difference. The relaxation rates $s_v$ and $\zeta_\alpha$ are determined by

$$s_v^{-1} = \frac{1}{2} + \frac{MaJL\sqrt{3Pr}}{c\delta_t\sqrt{Ra}}, \quad \zeta_\alpha^{-1} = \frac{1}{2} + \frac{\Gamma c_s^2\left(s_v^{-1} - 0.5\right)}{Jc_{sT}^2 Pr} \tag{46}$$

where $Ma = u_c/c_s$ is the Mach number ($u_c = \sqrt{g\beta\Delta TL}$ is the characteristic velocity). For incompressible thermal flows considered in this work, the Mach number $Ma$ is set to 0.1.

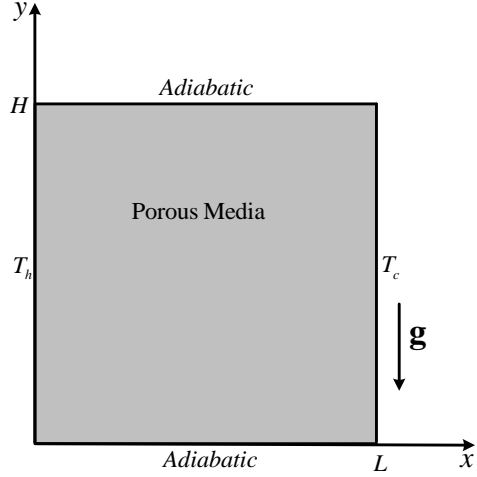

Fig. 5. Schematic diagram of natural convection in a porous cavity.

In simulations, we set $Pr = 1$, $J = \Gamma = \sigma = 1$, and the grid sizes of $128 \times 128$, $192 \times 192$, and $256 \times 256$ are employed for $Da = 10^{-2}$, $10^{-4}$, and $10^{-6}$, respectively. In Fig. 6, the streamlines and isotherms (the normalized temperature is $\theta = (T - T_0)/\Delta T$ with $T_0 = (T_h + T_c)/2$ being the reference temperature) for $Ra^* = 10^3$ ($Ra^* = RaDa$ is the Darcy-Rayleigh number) and $\phi = 0.6$ are illustrated. From Fig. 6 we can observe that as $Da$ decreases, the velocity and thermal boundary layers near the vertical walls become thinner. As $Da$ increases to $10^{-2}$, more convective mixing occurs inside the cavity and the isotherms are less crowded near the corners. The above observations indicate that for the same $Ra^*$, the heat transport in the Darcy regime ($Da = 10^{-6}$) is higher than that in the non-Darcy regime. The observed phenomena from the flow and temperature fields agree well with previous studies [34,49]. To quantify the results, the average Nusselt numbers of the hot wall are measured and listed in Table 1. The numerical results given by Nithiarasu et al. [49] using the finite-element method and the numerical results obtained by Guo and Zhao [34] using a BGK-LB method are also listed in Table 1 for comparison. It can be clearly seen that the present results agree well with the well-documented results in the literature [34,49].

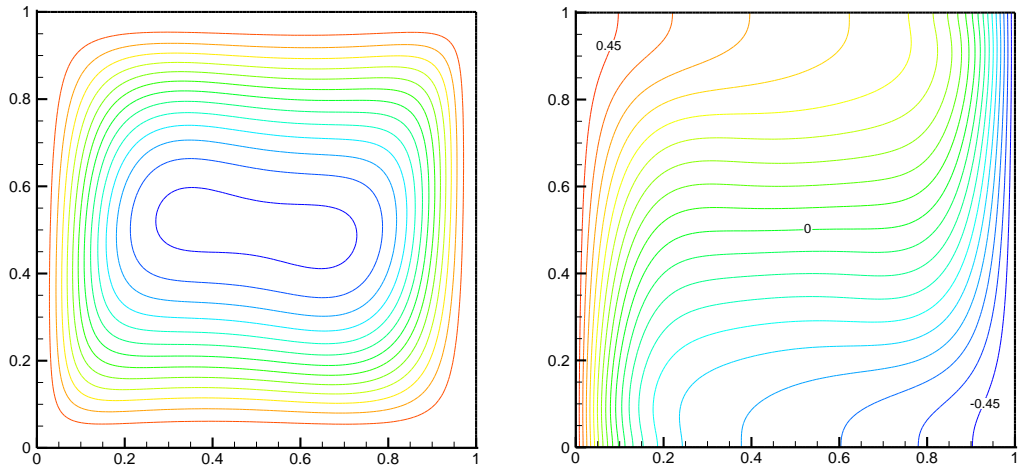

(a) $Da = 10^{-2}$, $Ra = 10^5$

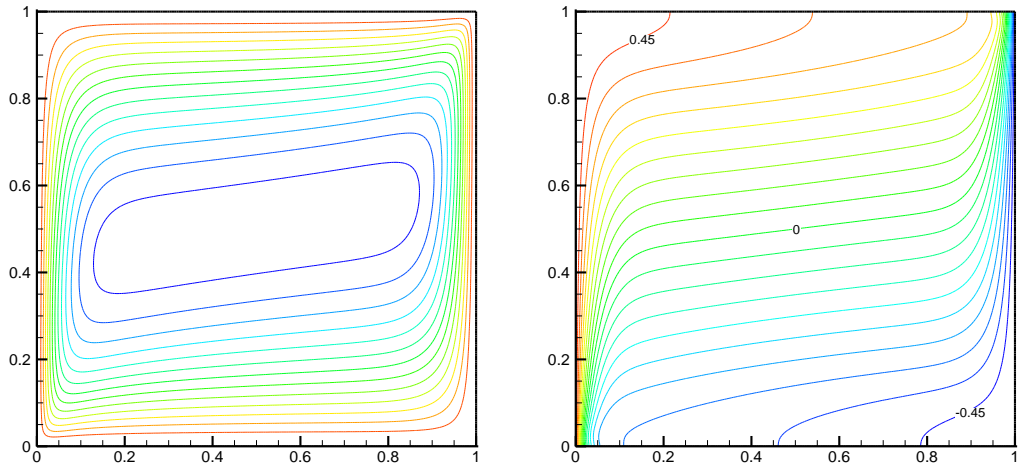

(b) $Da = 10^{-4}$, $Ra = 10^7$

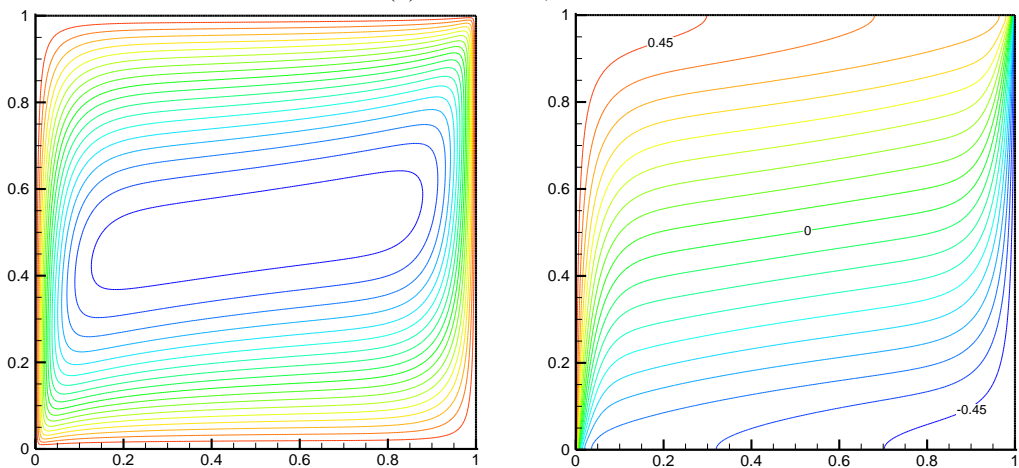

(c) $Da = 10^{-6}$, $Ra = 10^9$

Fig. 6. Streamlines (left) and isotherms (right) of natural convection in a porous cavity for $Ra^* = 10^3$ with $\phi = 0.6$.

Table 1 Comparison of the average Nusselt numbers with the values in Refs. [34,49] for various $Ra$, $\phi$, and $Da$ with $Pr = 1.0$.

| $Da$ | $Ra$ | $\phi = 0.4$ | | | $\phi = 0.6$ | | | $\phi = 0.9$ | | |
|---|---|---|---|---|---|---|---|---|---|---|
| | | Ref.[49] | Ref.[34] | Present | Ref.[49] | Ref.[34] | Present | Ref.[49] | Ref.[34] | Present |
| $10^{-2}$ | $10^3$ | 1.010 | 1.008 | 1.0078 | 1.015 | 1.012 | 1.0121 | 1.023 | - | 1.0176 |
| | $10^4$ | 1.408 | 1.367 | 1.3572 | 1.530 | 1.499 | 1.4873 | 1.640 | - | 1.6268 |
| | $10^5$ | 2.983 | 2.998 | 2.9988 | 3.555 | 3.422 | 3.4429 | 3.910 | - | 3.9248 |
| $10^{-4}$ | $10^5$ | 1.067 | 1.066 | 1.0642 | 1.071 | 1.068 | 1.0671 | 1.072 | - | 1.0693 |
| | $10^6$ | 2.550 | 2.603 | 2.6039 | 2.725 | 2.703 | 2.7118 | 2.740 | - | 2.7957 |
| | $10^7$ | 7.810 | 7.788 | 7.7828 | 8.183 | 8.419 | 8.4856 | 9.202 | - | 9.3121 |
| $10^{-6}$ | $10^7$ | 1.079 | 1.077 | 1.0758 | 1.079 | 1.077 | 1.0786 | 1.08 | - | 1.0810 |
| | $10^8$ | 2.970 | 2.955 | 2.9802 | 2.997 | 2.962 | 3.0156 | 3.00 | - | 3.0535 |
| | $10^9$ | 11.46 | 11.395 | 11.6642 | 11.79 | 11.594 | 12.020 | 12.01 | - | 12.1258 |

The effect of the heat capacity ratio $\sigma$ on the heat transfer process inside the porous cavity is investigated. Theoretically, for convection heat transfer in porous media under local thermal equilibrium condition, $\sigma$ has no influence on the solutions at the steady state (the transient term in Eq. (3) tends to 0 at the steady state). In Fig.7, the time history of the average Nusselt numbers of the hot wall for different values of $\sigma$ with $Ra=10^5$, $Da=10^{-2}$, and $\phi=0.6$ are plotted. As shown in the figure, $\sigma$ only has influence on the heat transfer process at the unsteady state, and a smaller $\sigma$ makes a shorter response time (the temperature field reaches the steady state more quickly). At the steady state, the average Nusselt numbers for different values of $\sigma$ are the same ($\overline{Nu} = 3.4429$).

In what follows, the computational efficiency of the present CLB method is studied. In Table 2, the computational times of different LB methods for $Ra = 10^7$, $Da=10^{-4}$, and $\phi = 0.6$ are presented. Here, the (non-orthogonal) MRT-LB method is obtained by setting $\mathbf{N}$ and $\mathbf{N}_T$ to unit matrices, and the equilibrium distribution functions of the BGK-LB method are obtained via $\left| f_i^{eq} \right\rangle = \mathbf{M}^{-1} \left| m_i^{eq} \right\rangle$ and $\left| g_i^{eq} \right\rangle = \mathbf{M}_T^{-1} \left| n_i^{eq} \right\rangle$. As shown in the table, the computational time of the MRT-LB method is about 15% more than that of the BGK-LB method, while the CLB method is about 33% slower than the BGK-LB counterparts. In comparison with the MRT-LB method, computational procedures related with the shift matrices are needed in the CLB method, hence more computational time is required in simulations.

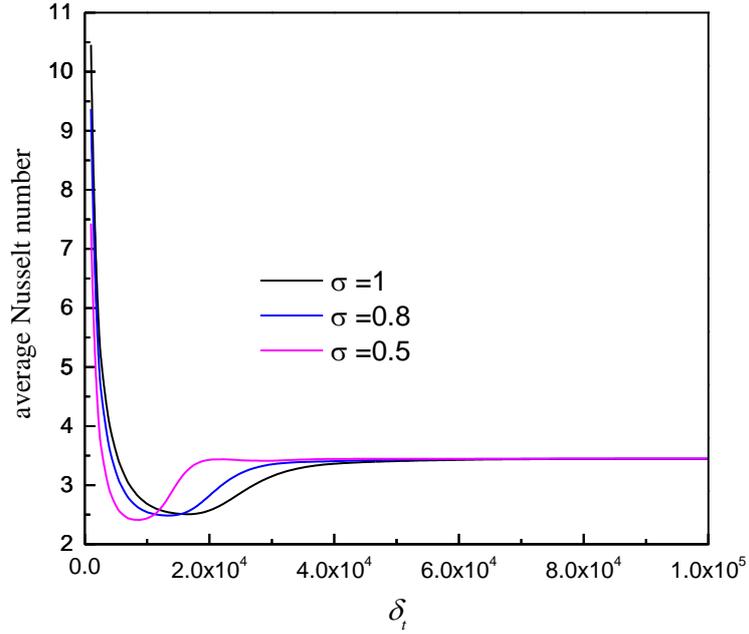

Fig.7. The time history of the average Nusselt numbers of the hot wall for different values of $\sigma$ with $Ra=10^5$, $Da=10^{-2}$, and $\phi=0.6$.

Table 2 Comparisons of the computational times of the CLB method with those of the MRT-LB and BGK-LB methods.

| Steps | Method | CPU time (s) | $t_{cpu}/t_{cpu,BGK}$ |
|---|---|---|---|
| $2\times10^5$ | BGK | 1568.92 | 1.0 |
|  | MRT | 1812.31 | 1.155 |
|  | Present | 2099.33 | 1.338 |
| $5\times10^5$ | BGK | 3996.87 | 1.0 |
|  | MRT | 4578.45 | 1.146 |
|  | Present | 5321.94 | 1.332 |

## 4. Conclusions

In this paper, an REV-scale CLB method for convection heat transfer in porous media is presented. In the CLB method, the influence of the porous media is considered by introducing the porosity and heat capacity ratio into the shift matrices. As a result of this strategy, the practical implementation of the present REV-scale CLB method is simple. Numerical simulations of mixed convection flow in a porous channel and natural convection in a porous cavity are carried out to demonstrate the practicability and accuracy of the present CLB method. Numerical results indicate that the present CLB method can serve as an efficient and powerful numerical tool for large-scale engineering calculations of convection heat transfer in porous media. Extensions of the present CLB method to simulate solid-liquid phase-change heat transfer in porous media will be considered in future studies.

**Acknowledgements**

The authors gratefully acknowledge the Initial Scientific Research Fund for High-level Talents (1608719045), the Initial Scientific Research Fund for Special Zone's Talents (XJ18T06), and the Scientific Research Program Funded by Shaanxi Province Education Department (19JK0905).

**Appendix A: The inverse matrices of $\mathbf{N}$ and $\mathbf{N}_T$**

$$\mathbf{N}^{-1} = \begin{bmatrix} 1 & 0 & 0 & 0 & 0 & 0 & 0 & 0 & 0 \\ u_x & 1 & 0 & 0 & 0 & 0 & 0 & 0 & 0 \\ u_y & 0 & 1 & 0 & 0 & 0 & 0 & 0 & 0 \\ u_x^2/\phi & 2u_x/\phi & 0 & 1 & 0 & 0 & 0 & 0 & 0 \\ u_y^2/\phi & 0 & 2u_y/\phi & 0 & 1 & 0 & 0 & 0 & 0 \\ u_xu_y/\phi & u_y/\phi & u_x/\phi & 0 & 0 & 1 & 0 & 0 & 0 \\ u_x^2 u_y/\phi & 2u_xu_y/\phi & u_x^2/\phi & u_y & 0 & 2u_x & 1 & 0 & 0 \\ u_y^2 u_x/\phi & u_y^2/\phi & 2u_xu_y/\phi & 0 & u_x & 2u_y & 0 & 1 & 0 \\ u_x^2 u_y^2/\phi^2 & 2u_xu_y^2/\phi^2 & 2u_yu_x^2/\phi^2 & u_y^2/\phi & u_x^2/\phi & 4u_xu_y/\phi & 2u_y/\phi & 2u_x/\phi & 1 \end{bmatrix} \quad (A1)$$

$$\mathbf{N}_T^{-1} = \begin{bmatrix} 1 & 0 & 0 & 0 & 0 \\ u_x/\sigma & 1 & 0 & 0 & 0 \\ u_y/\sigma & 0 & 1 & 0 & 0 \\ u_x^2/\sigma & 2u_x & 0 & 1 & 0 \\ u_y^2/\sigma & 0 & 2u_y & 0 & 1 \end{bmatrix} \quad (A2)$$